# Fractal Art Generation using GPUs

W.D. Mayfield[a], J.C. Eiland[a], T.J. Hutyra[a], M.C. Paulsen[a], and B.M. Wyatt[a]

[a] *Department of Mathematics, Tarleton State University, Stephenville, Texas, USA*

**Abstract**

Fractal image generation algorithms exhibit extreme parallelizability. Using graphics processing units (GPUs) to implement parallel escape-time algorithms for Julia sets of functions, we produce visually attractive fractal images much faster than traditional serial methods. Vastly improved speeds allow real-time generation and display of images. A comparison is made between sequential and parallel implementations of the algorithm. An application created by the authors demonstrates using the increased speed to create dynamic imaging of fractals where the user may explore paths of parameter values corresponding to a given function's Mandelbrot set. Examples are given of artistic and mathematical insights gained by experiencing fractals interactively and from the ability to sample the parameter space quickly and comprehensively.

## 1 Introduction

In 'The Postmodern Beauty of Fractals,' Garousi explains the procedures of a fractal artist. First, an artist selects the computational aids required for fractal image generation. These tools employ algorithms that allow the artist to explore the different components of fractal image creation. Next, there is the selection of some mathematical function that has been shown to have fractal properties; that is, it has self-similarity, and in general, a degree of complexity and often chaos that allows for an infinite number of possible variations. Finally, after choosing the specifics of mathematical function and algorithm, the artist must select avenues of visual representation for color and form to take final shape.

It is in this selection of function and algorithm that the artist most looks to the innovations of mathematics; efficiency and variety of technology follow from the work of computer scientists. While [8] has discussed deeper, more creative variations on the algorithms that artists and programmers employ in fractal art, here we focus on how new computational tools inform selection of specific mathematical functions. The tool explored is the general purpose graphics processing unit (GPU)—its effect is that, through a dramatic increase in speed, artists may dynamically modify the parameters which determine function behavior and, ultimately, the image created.

The rest of Garousi's essay describes the postmodern nature of fractal art [4]. One of the hallmarks of postmodernity is living in the face of *complexity*, something the fractal artist encounters at every turn. Fractal art, with its own infinite complexities, falls in line very well with the idea of postmodernity. This art allows fractal geometry, already one of the closest encounters of pure mathematics with the inner workings of nature (see [1], [5]) to speak to humans facing the unpredictability of chaos and complexity. Garousi sees this in the creative process as well, where artists experience a 'random playing around' within an infinitely vast parameter space.

Yet the age of postmodernism continues to evolve, and has recently incorporated itself with what we have come to call the era of big data. Both of these concepts react to the vastness and complexity of existence, but while postmodernity involves a level of hopelessness, big data—without claiming to find *ultimate* truths—attempts to at the very least get a *handle* on the world we felt was unfathomable. Big data uses technological muscle and exploits parallelism to glean as much insight as possible from the profusion of data that surrounds us [6]. For fractal artists, this is analogous to how the GPU so efficiently distributes the work of image generation that we can traverse parameter spaces in ways that were not possible only a few years ago.

We use the speed of parallel processing to allow generation of fractal images in real-time as parameters are varied. Not only does this open up the world of dynamic fractal art, but it allows fractal artists to traverse vast parameter spaces much faster, so that functions can be vetted for aesthetic possibilities without reliance on guesswork.

## 2 Sequential vs. parallel implementations

Here we focus on creating images of Julia sets of functions using escape-time algorithms. These algorithms involve an iterative function run on values in the complex plane, represented by the computer as locations of pixels on the screen. We will use the function $Z_{n+1} = Z^2 + C$, where $C$ is a complex constant, though many other functions yield fruitful explorations of fractal art. Each initial value $Z_0$ is selected via what [8] calls the 'region-covering routine,' where pixels are scaled to the complex plane and are assigned color levels according to the number of iterations that initial value takes to either diverge or not under the given function. A limit is placed on the number of times the function is iterated; in our implementation, that limit is 100. GPU architecture is particularly well-suited to this 'embarrassingly parallel' task, and allows for incredible speed increases over traditional sequential algorithms run on central processing units (CPUs) [7].

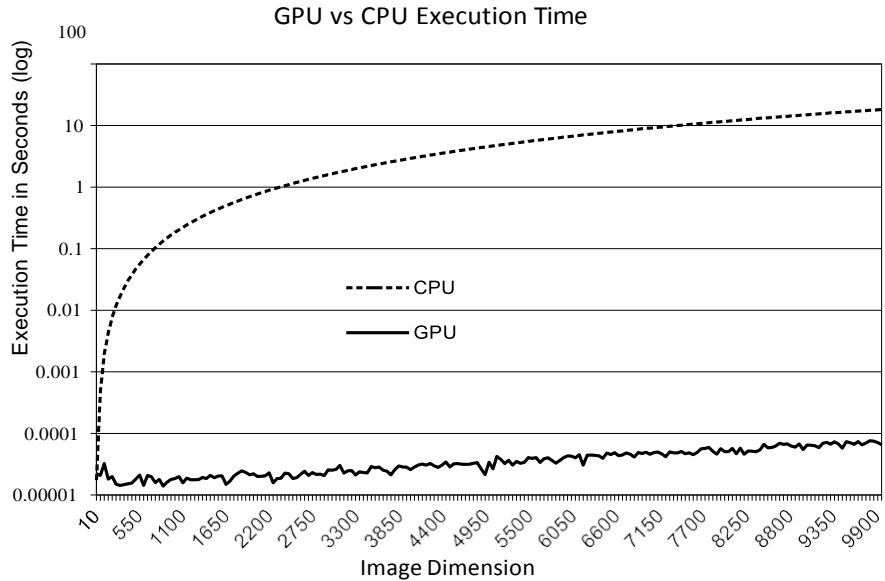

Figure 1: Log scale comparison of CPU and GPU execution times with generic implementations of a Julia set escape-time algorithm.

In order to appreciate how GPUs have revolutionized the way these images are produced, we implemented both sequential and parallel versions of a generic escape-time algorithm. The sequential version uses a main loop to cover the pixel region, while GPUs virtually assign processors to each point. In most other ways the two algorithms behave identically. See [7] for an example of a parallel implementation for this algorithm. The C programming language was used, with the compute unified device architecture (CUDA) extension for GPU programming. We did not include the time each program spent for displaying the actual images, so as to compare only the algorithm's raw computation times. This speed comparison used an Intel Core i7-4770K CPU and an NVIDIA GeForce GTX TITAN graphics card. We varied image dimension (always square) as a way of comparing the two methods, from as small as 10-by-10 pixels to 10,000-by-10,000 pixels. See Figure 1 for a log-scaled plot of generation times for both methods.

In Figure 1, we see that for images even a few hundred pixels across, the sequential algorithm's generation time increases to the point that even without the overhead of image display, on-the-fly rendering would not be possible, as the frames would update far too slowly to achieve smooth, dynamic fractal art, and the resolution of parameters sampled would be limited. On the other hand, the massive parallelism of the GPU performs very well, such that computation time increases much more slowly when we increase image resolution. In our main implementation for dynamic fractal generation, full-screen images were produced and displayed at over 50 frames per second.

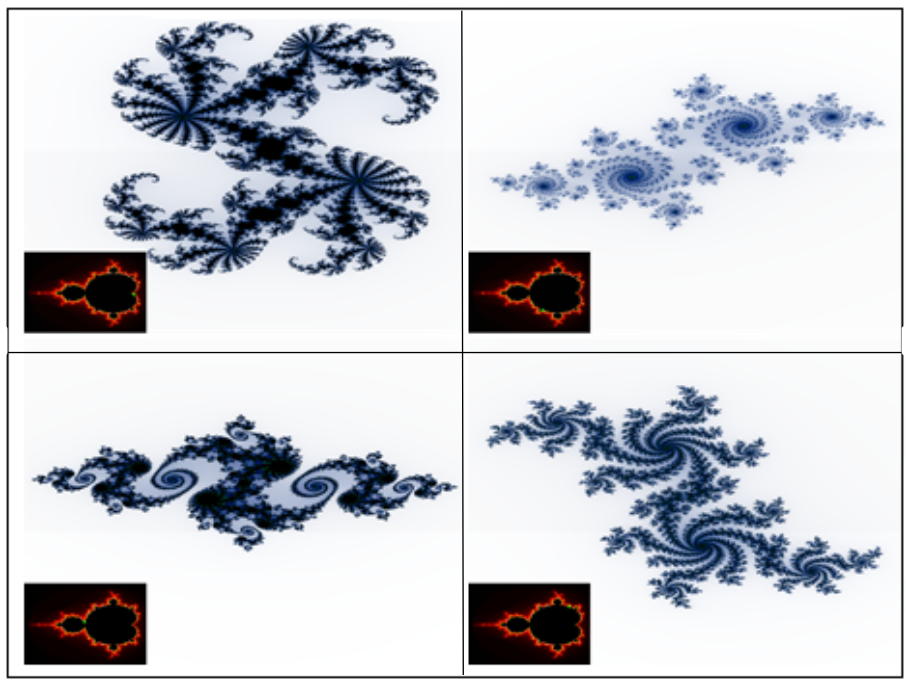

Figure 2: Screenshots showing the Julia sets generated by parameter values traversing the main cardioid of the Mandelbrot set. Clockwise from top-left, the parameter values are $C = .320564 − .0391827i$, $C = −.454038 − .572187i$, $C = −.763667 + .0870413i$, and $C = .137384 + .600803i$.

## 3  Traversing the fractal parameter space

Our implementation of dynamic fractal image generation places emphasis on the ability to traverse paths of parameter values. The parameter in question is the constant $C$ in the function given above. [1] provides many useful results that our program exploits. First, slight variations in parameter values correspond to smooth transitions in the shape of the Julia set. This is essential for the creation of dynamic fractal art. Also of importance is the Mandelbrot set for the same function used to generate the Julia set. The Mandelbrot set is generated by choosing values for $C$ from the complex plane (corresponding to pixels), instead of $Z_0$ as in the Julia set—in the Mandelbrot set, $Z_0$ is given the value 0. Barnsley presents the Mandelbrot set for a function as a roadmap for parameter values to generate the Julia set, where values near the boundary of the Mandelbrot correspond to the most complexity in Julia sets. This is verified visually when using fractal image generation programs.

To control parameter values for the constant $C$ in our implementation, we display a miniature view of the Mandelbrot set for $Z^2 + C$ in the lower left corner of the screen, and place a green cursor at the current value of the parameter. The user may vary the parameter value via arrow keys and mouse clicks. As the parameter value moves across various features of the Mandelbrot set, corresponding changes in the Julia set are displayed, so that the viewer experiences the fractal dynamically.

The user may also choose to have the parameter move through prescribed paths. Figure 2 depicts four screen captures as the parameter moves clockwise along the cardioid given by $f(t) = ([2\cos t - \cos 2t]/a, [2\sin t - \sin 2t]/a)$, where $a$ has the value 3.9. The border of the main cardioid of the Mandelbrot set is this function when $a = 4$, so using a slightly smaller value for $a$ assures that the value is always taken from just outside the main body of the Mandelbrot set. Traversing this path allows us to see that the main plane of the Julia set rotates by 180 degrees as the parameter moves around the Mandelbrot set's boundary. Then, if we let the size of the cardioid increase gradually (by decreasing the values of $a$ above), almost all relevant parameter values at this scale will be traversed after a few trips around the Mandelbrot set. For a fractal artist, or perhaps a graphic designer in need of fractal images with specific features and orientation, this information is invaluable.

One way fractal artists exploit the images created by these functions is by examining deep zooms on portions of the fractal. This program not only allows deep zooms on the Julia set, but also on the Mandelbrot set, so that parameters may be varied with extreme precision among the Mandelbrot set's features. Figure 3 depicts such a zoom, and allows us to see how even minute variations can affect the main features of the corresponding Julia set.

## 4  Discussion

The exploration of parameter spaces in real-time goes far beyond the ability to hone in on targeted aesthetic images of fractals. The visualization of the mathematics behind fractal generation is fascinating, which lends itself very well to educational purposes. Inaccessible results become obvious when we are given the ability to freely experience the effects of varying parameters. For example, the bulbs attached to the main cardioid of the Mandelbrot set correspond to convergence on varying-degree orbits of those values under iteration, where the degree increases as the size of the bulbs decrease. When passing through these parameter values, we see the degree of orbit show up in features of the corresponding Julia set (this can be seen in Figure 2). These types of insights can be explored for mathematical and aesthetic purposes, as users can interact with the mathematics of chaos without the need for rigorous background work.  In [9], we see how at all levels of education the interest piqued by such explorations is extremely valuable.

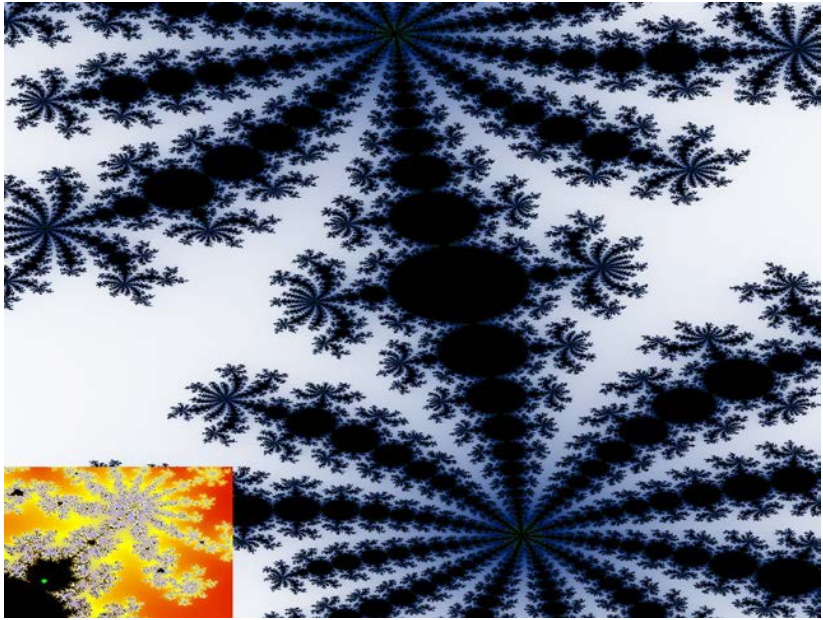

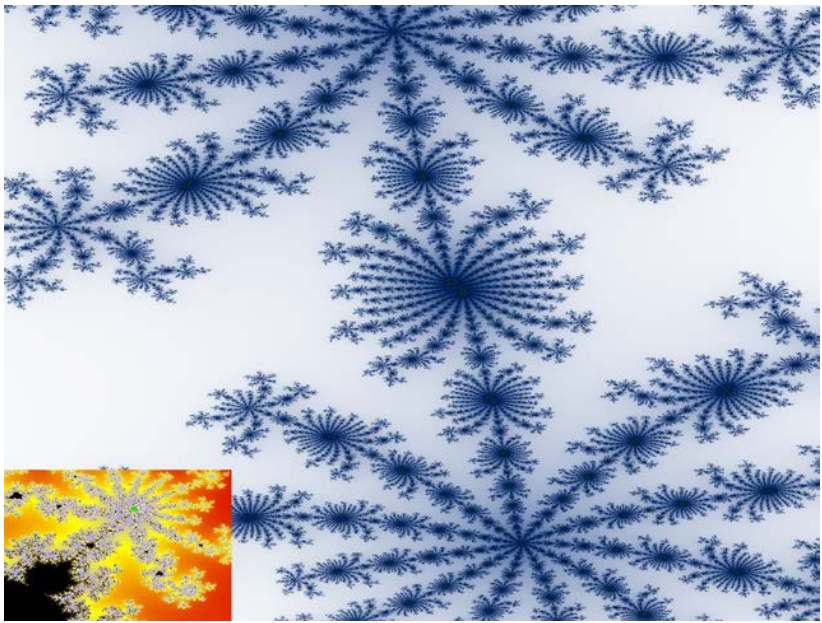

Figure 3: Screenshots of Julia set features for two different but very close parameter values, $C = .177078 + .577384i$ and $C = .185723 + .588104i$ taken from a deep zoom on the Mandelbrot set.

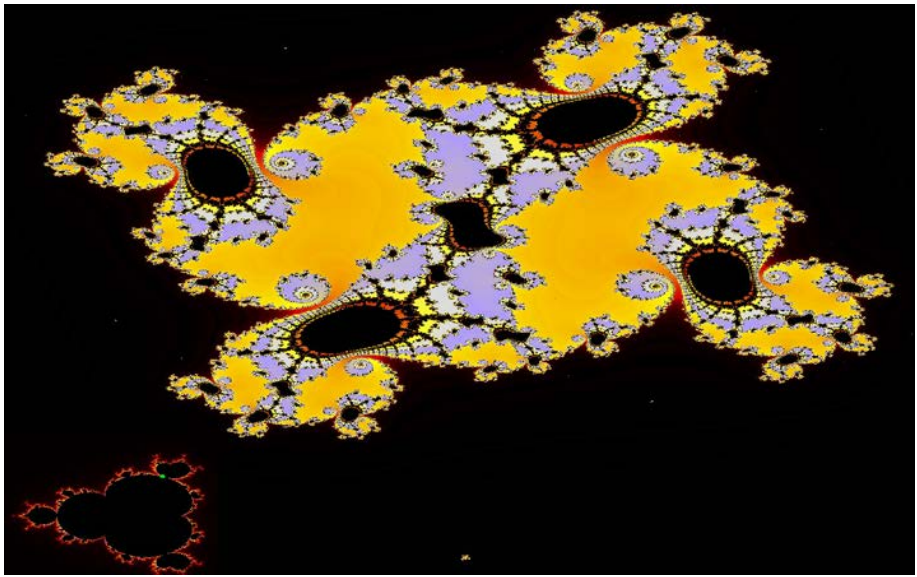

Figure 4: Screenshot of zoom on Julia set for function $z^4 + (z^2 + 1)/(z^2 + 1) + C$ and parameter value $C = .862085 + .64695i$ using another coloring scheme.

The implications of GPU programming for fractal mathematics also reach into more scientific studies. Fractals, while they have not permeated the main-stream of mathematical modeling, can be used to represent phenomena in ways that Euclidean mathematics cannot. Fractal interpolation has been explored, and is likely just as parallelizable as image generation ([10]). One of the most important tasks in fractal modeling is determining fractal dimension—often computationally intensive, but which has been shown to be a highly parallel process ([2]). Mathematical modeling is already limited by computational ability as we press for more and more precision, and fractal functions invariably introduce additional complexity. Yet the approach lends itself well to parallel programming, where the GPU may eventually pave the road to models that more accurately represent natural phenomena.

## 5 Conclusion

In one of the more famous intersections of mathematics and art—in this case, literature—Tom Stoppard's work *Arcadia* explores the idea of iterative functions representing mechanisms that lie deep within the natural world. '...an algorithm is a recipe, that if you knew the recipe to produce a leaf, you could then easily iterate the algorithm to draw a picture of the leaf' ([3]). The hard part, then, is discovering such a recipe from the infinite space of functions and parameters that might describe such phenomena. Finding the right functional form and parameters, though, means a colossal search of trial and error, corrections made, leads followed, revision and refinement, until eventually we discover the facsimile of nature in pure mathematics. The search is worth the effort.

Our program makes a venture into the possibilities describes above. Future work plans to make an exploration of multiple parameters in more complex functions possible, and to allow more fluid modification of functional form (see Figure 4 for one variation). Also, we continually look for more accessible visualizations of the mathematics within fractals.